\documentclass{PoS}

\usepackage{rotating}
\usepackage{mathrsfs}

\newcommand{\sqrtsnn}{\sqrt{s_{_{\ensuremath{\it{NN}}}}}}
\newcommand\qhat{\hat{q}}
\newcommand{\jpsi}{J/\psi}

\newcommand{\ups}{\Upsilon}
\newcommand{\upsp}{\Upsilon^{'}}
\newcommand{\upspp}{\Upsilon^{''}}
\newcommand{\ecrit}{\varepsilon_{\mbox{\tiny{\rm crit}}}}
\newcommand{\Tcrit}{T_{\mbox{\tiny{\rm crit}}}}

\title{Perturbative probes of QCD matter at the LHC}

\ShortTitle{Perturbative probes of QCD matter at the Large Hadron Collider}

\author{\speaker{David d'Enterria}\\ 
        CERN, PH Department, 1211 Geneva, Switzerland\\
        E-mail: \email{david.d'enterria@cern.ch}}

\abstract{
The main results on electroweak probes, jets, high-$p_T$ hadrons, heavy-flavour and quarkonia production 
from the first two years of heavy-ion operation at the Large Hadron Collider (LHC) are briefly reviewed. 
Data measured at center-of-mass energies $\sqrtsnn$~=~2.76~TeV in lead-lead (Pb-Pb) collisions are compared 
to proton-proton (p-p) measurements in order to extract
information on the properties of hot and dense strongly-interacting matter.}

\FullConference{6th International Conference on Quarks and Nuclear Physics\\
		April 16-20, 2012\\
		Ecole Polytechnique, Palaiseau,  Paris}

\begin{document}

\section{Introduction}

Nucleus-nucleus collisions (A-A) at ultrarelativistic energies provide the experimental means to study
the (thermo)dynamics of quarks and gluons under extreme conditions of temperature and density.
Head-on collisions of Pb ions at LHC energies produce very hot and dense matter by 
concentrating a substantial amount of energy $\mathscr{O}$(2 TeV) in an extended volume 
$\mathscr{O}$(150~fm$^3$) at thermalisation times of $\tau_0$~=~1~fm/c~\cite{Chatrchyan:2012mb}.
Such energy densities are more than one order of magnitude above the critical value,
$\ecrit\approx$~1~GeV/fm$^3$, predicted by lattice quantum chromodynamics (QCD)
calculations~\cite{Karsch:2001cy} for the formation of a deconfined system of bare-mass quarks and gluons
(Quark Gluon Plasma, QGP)~\cite{shuryak77}.
\begin{figure}[htpb]
\begin{minipage}{0.52\linewidth}
\centering
\includegraphics[width=0.99\columnwidth, bb = 0 58 561 660,clip=true]{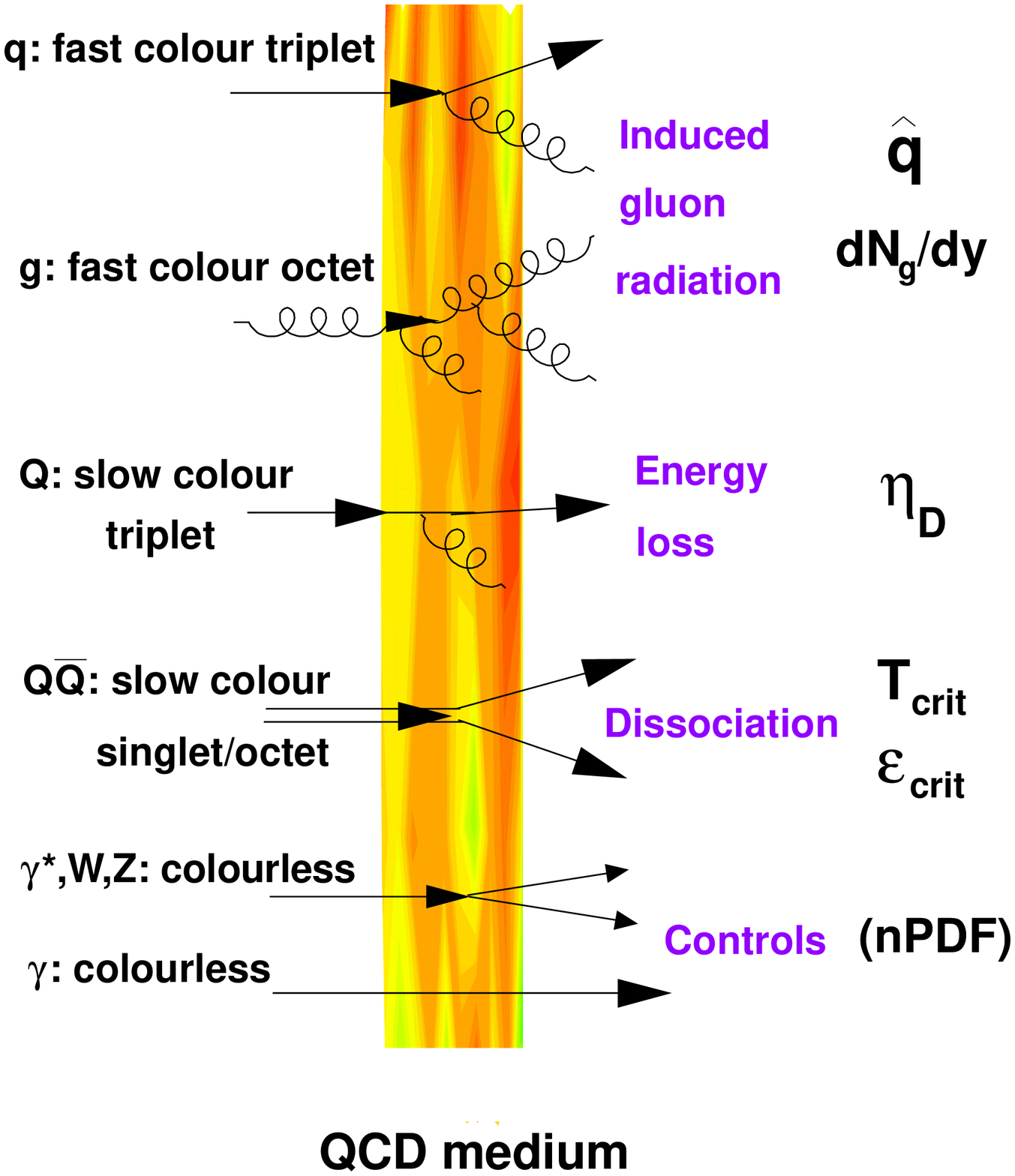}
\end{minipage}
\begin{minipage}{0.48\linewidth}
\centering
\includegraphics[width=0.90\columnwidth,height=4.cm,clip=true]{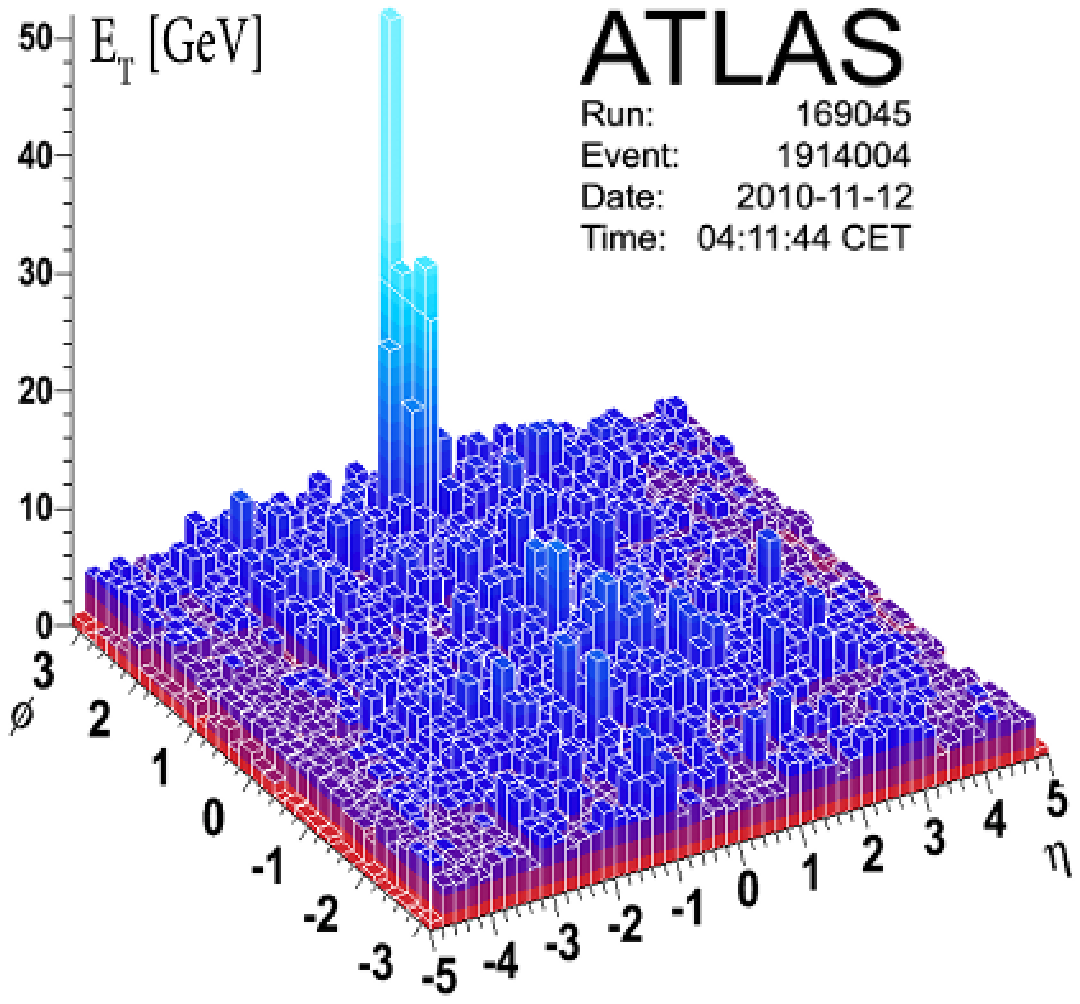}
\includegraphics[width=0.88\columnwidth,height=4.2cm,clip=true]{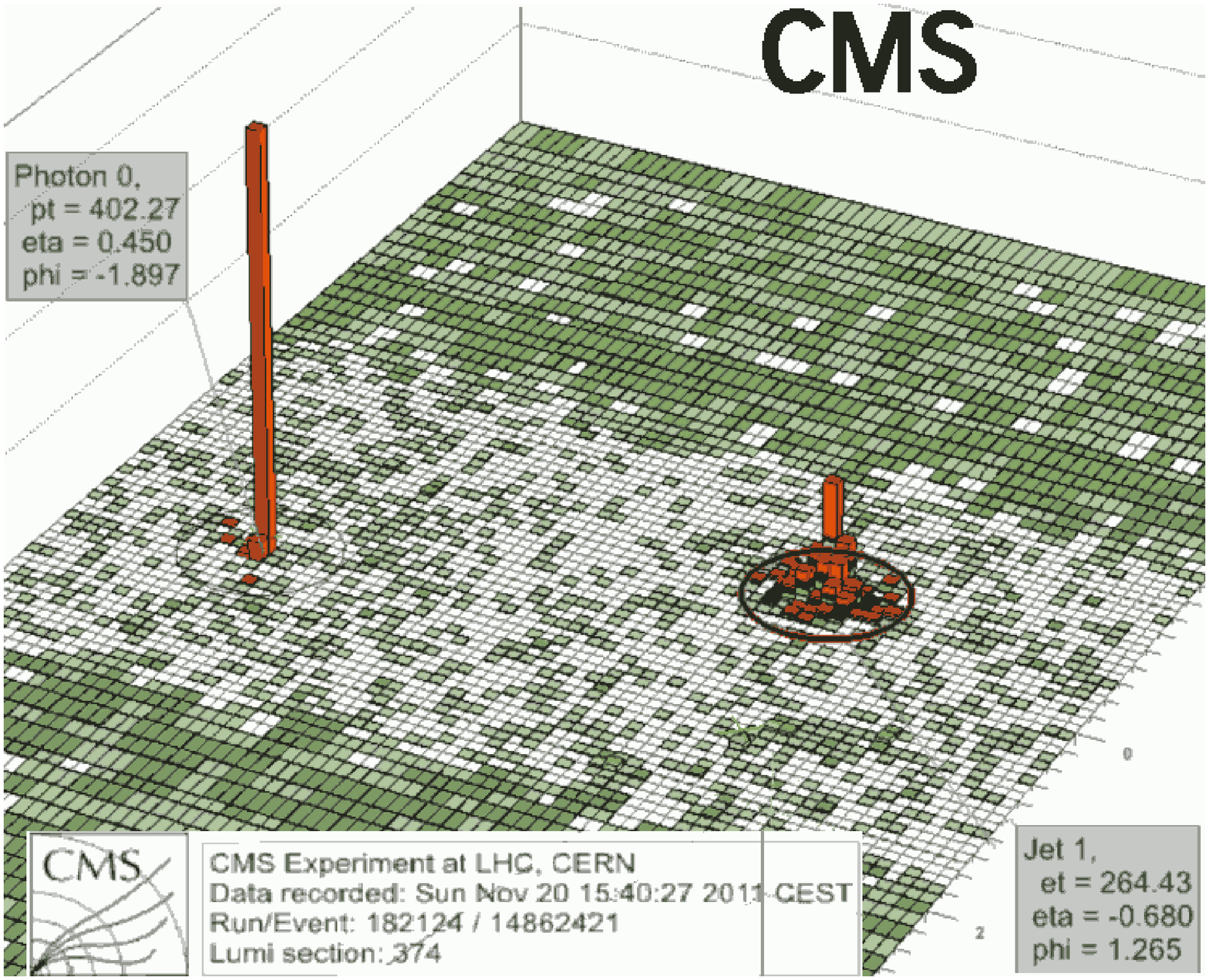}
\end{minipage}%
\caption{Left: Examples of the sensitivity of various perturbative probes to quark-gluon matter properties in A-A collisions~\cite{d'Enterria:2006su}.
Right: ``Jet quenching'' event displays in central Pb-Pb collisions at the LHC: 
monojet-like event~\cite{Aad:2010bu} (top) and $\gamma$-jet event 
with a recoiling jet with reduced energy~\cite{Chatrchyan:2012gt} (bottom).}
\label{fig:qgp_probes}
\end{figure}
Among all experimental observables, particles with large transverse momentum $p_T$ and/or high mass $m$
(``hard probes'') are useful tomographic tools of the produced medium (Fig.~\ref{fig:qgp_probes} left) 
as: (i) they originate from partonic scatterings with large momentum transfer $Q^2$ 
and couple directly to the fundamental quark and gluon degrees of freedom; 
(ii) their production time-scale is very short, $\tau\approx 1/(p_T,m)\lesssim$~0.1~fm/c, 
allowing them to propagate through (and be potentially affected by) the medium, and
(iii) their cross sections can be theoretically computed in perturbative QCD calculations
as a convolution of parton distribution (PDFs, $f_{a/A}$) and fragmentation (FFs, $D_{c\rightarrow h}$)
functions 
times the subprocess parton-parton scattering cross section:
$d\sigma^{hard}_{AB\rightarrow h}= f_{a/A}(x,Q^2)\otimes f_{b/B}(x,Q^2)\otimes 
d\sigma_{ab\rightarrow c}^{hard} \otimes D_{c\rightarrow h}(z,Q^2)$. 
In A-A collisions in the absence of final-state effects, the parton flux in a nucleus $A$ is the same as that
of a superposition of $A$ independent nucleons, $f_{a/A}\approx A\cdot f_{a/N}$, and thus
$d\sigma^{hard}_{AA\rightarrow h} \approx A^2 \cdot f_{a/N}(x,Q^2)\otimes \;f_{a/N}(x,Q^2)\otimes 
d\sigma_{ab\rightarrow c}^{hard}\otimes D_{c\rightarrow h}(z,Q^2)$.
The standard method to quantify the effects of the medium on a given hard probe
is via the ratio of A-A yields over p-p cross sections (scaled by the nuclear overlap function 
$T_{AA}(b)$ at impact parameter $b$): 
$R_{AA}(p_{T},y;b)\,=\,\frac{d^2N_{AA}/dy dp_{T}}{\langle T_{AA}(b)\rangle\,\times\, d^2 \sigma_{pp}/dy dp_{T}},$
which measures the deviation of A-A at $b$ from an incoherent superposition of nucleon-nucleon 
collisions. The study of the suppression (or enhancement) factors as a function of various variables provides
information on the medium transport coefficients, its temperature, energy density,
etc. (Fig.~\ref{fig:qgp_probes}, left).

\section{Unsuppressed electro-weak particle production}

\begin{figure}[htpb]
\includegraphics[width=0.52\columnwidth,height=5.cm,bb = 255 250 820 570,clip=true]{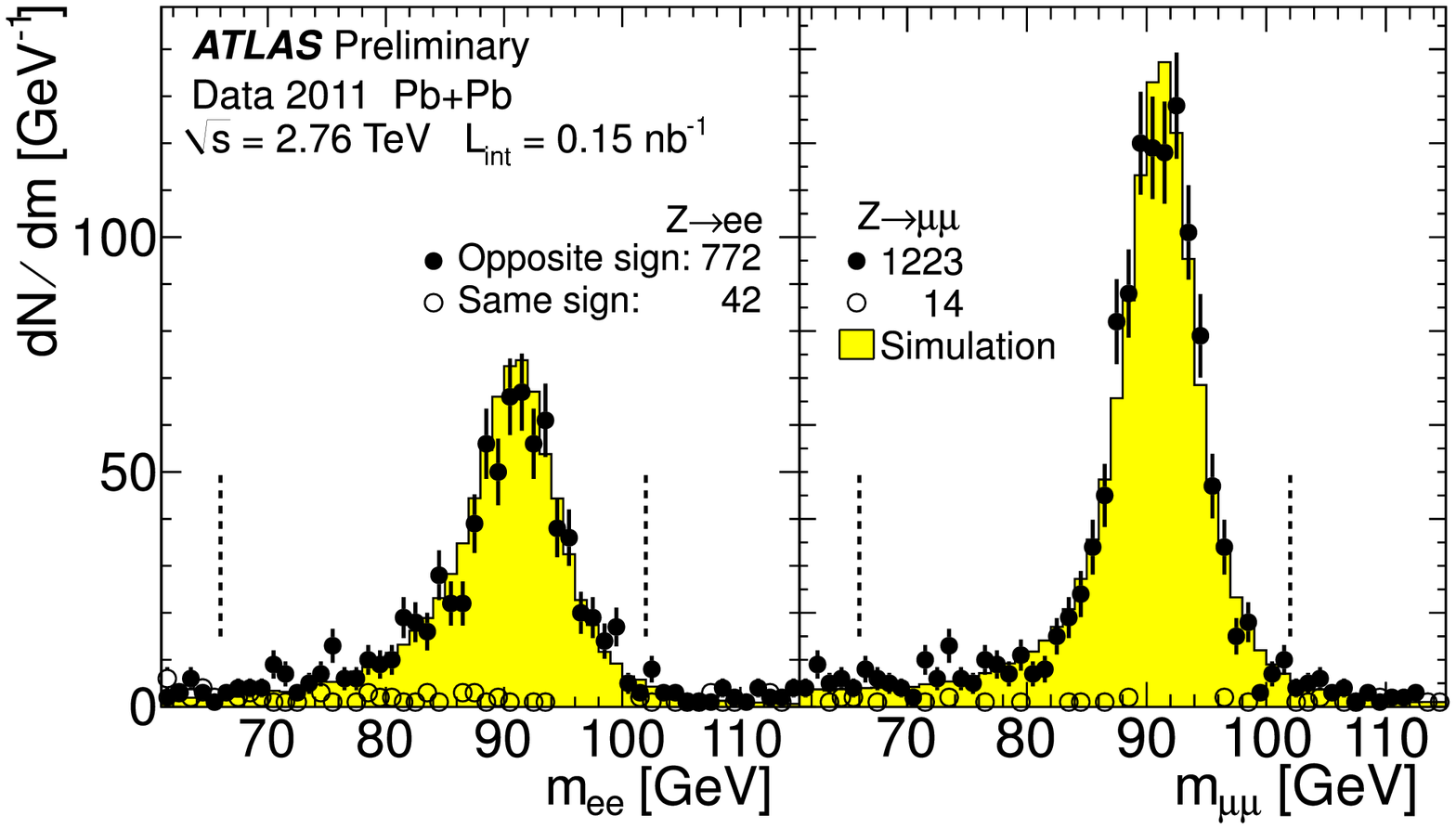}
\includegraphics[width=0.48\columnwidth,height=5.cm,clip=true]{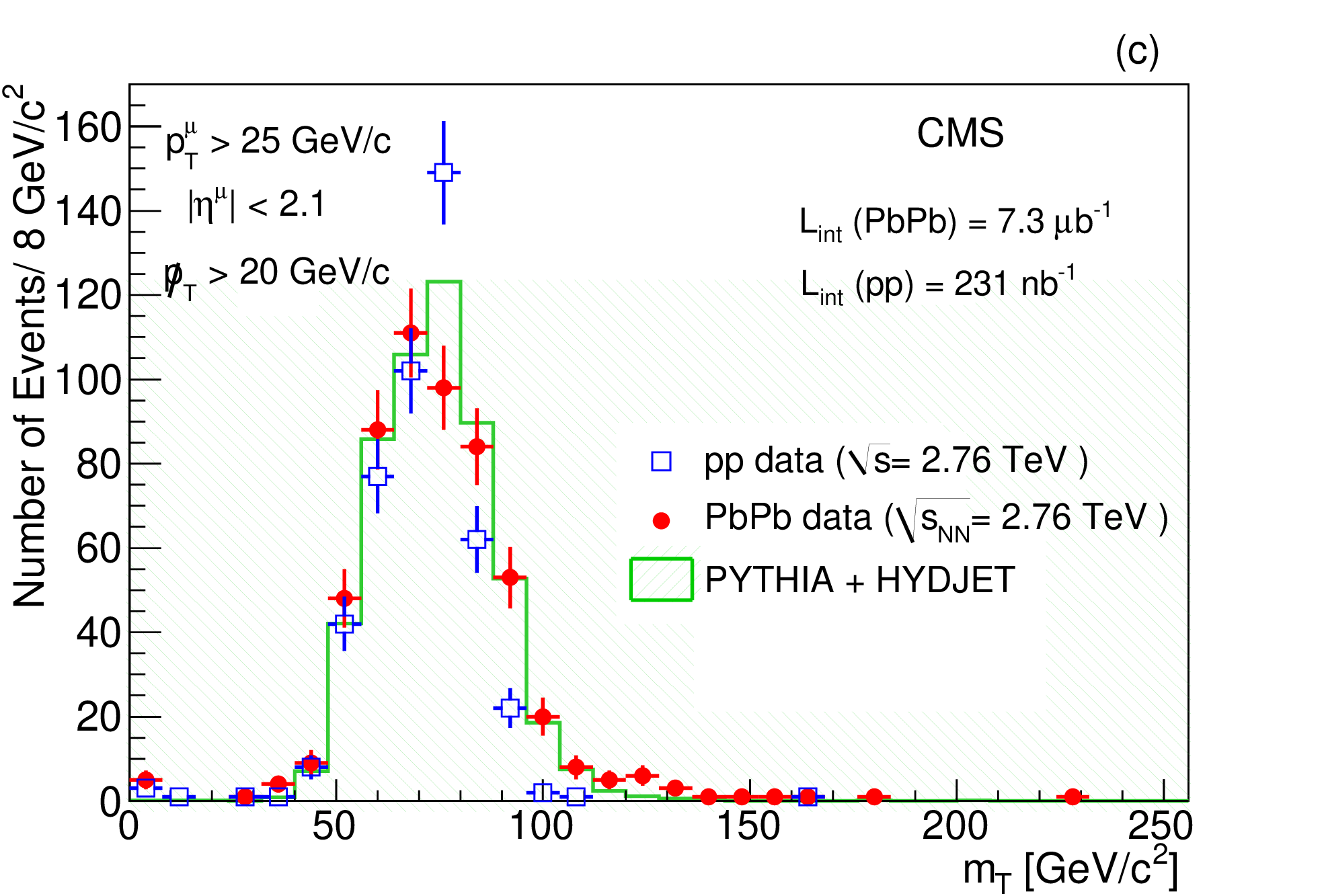}
\caption{Measured mass distributions for Z bosons ($\mu^+\mu^-$ and $e^+e^-$ invariant masses, left)~\cite{atlas:Z12} 
and W (transverse mass, right)~\cite{Chatrchyan:2012nt} in Pb-Pb (and p-p) collisions at 2.76 TeV.}
\label{fig:WZ}
\end{figure}

Weakly-interacting particles such as isolated photons, W, Z, and Drell-Yan pairs -- whose rates
are unaffected by final-state interactions in the produced medium -- are 
valuable benchmark processes in A-A collisions at collider energies. On the one hand, they allow one 
to experimentally confirm the validity of the perturbative ($A^2$) scaling of the p-p cross sections and,
on the other, they provide constraints on the nuclear PDFs which, in particular for the Pb case, 
are barely known from deep-inelastic e-A data~\cite{Eskola:2009uj}.
At the LHC, prompt-$\gamma$ have been measured above 20~GeV/c applying isolation cuts on background-subtracted
Pb-Pb collisions~\cite{Chatrchyan:2012vq}, Z bosons have been measured in the
dimuon~\cite{Aad:2010aa,Chatrchyan:2011ua} and dielectron~\cite{atlas:Z12} decay modes, and W bosons in
events with a high-$p_T$ muon and large missing transverse momentum from the undetected $\nu$~\cite{Chatrchyan:2012nt} (Fig.~\ref{fig:WZ}).
In all three cases, the measured yields for all Pb-Pb centralities are consistent with the corresponding p-p
cross sections scaled by the nuclear overlap function ($R_{AA}\approx$~1, Fig.~\ref{fig:worldRAA} left). 
Comparisons to NLO pQCD calculations indicate small nuclear PDF modifications as expected in the ranges
of parton fractional momentum $x$ and energy scale $Q^2$ probed~\cite{Eskola:2009uj}.
Interestingly, inclusive W production shows $R_{AA}$(W)~=~1.04~$\pm$~0.07~$\pm$~0.12 while individual
W$^\pm$ yields show differences due to isospin effects (unequal $u$ and $d$-quark content in Pb and p): 
$R_{AA}(\bar u d \to $W$^-)$~=~1.46 and $R_{AA}(u\bar d \to $W$^+)$~=~0.8~\cite{Chatrchyan:2012nt}. 
Future analyses of larger $\gamma,\gamma^\star$, W, Z data samples with reduced uncertainties, 
will provide enhanced constraints on the parton densities in nuclei.

\section{Suppression of high-$p_T$ hadrons}

\begin{figure}[htpb]
\includegraphics[width=0.60\columnwidth,height=6.2cm]{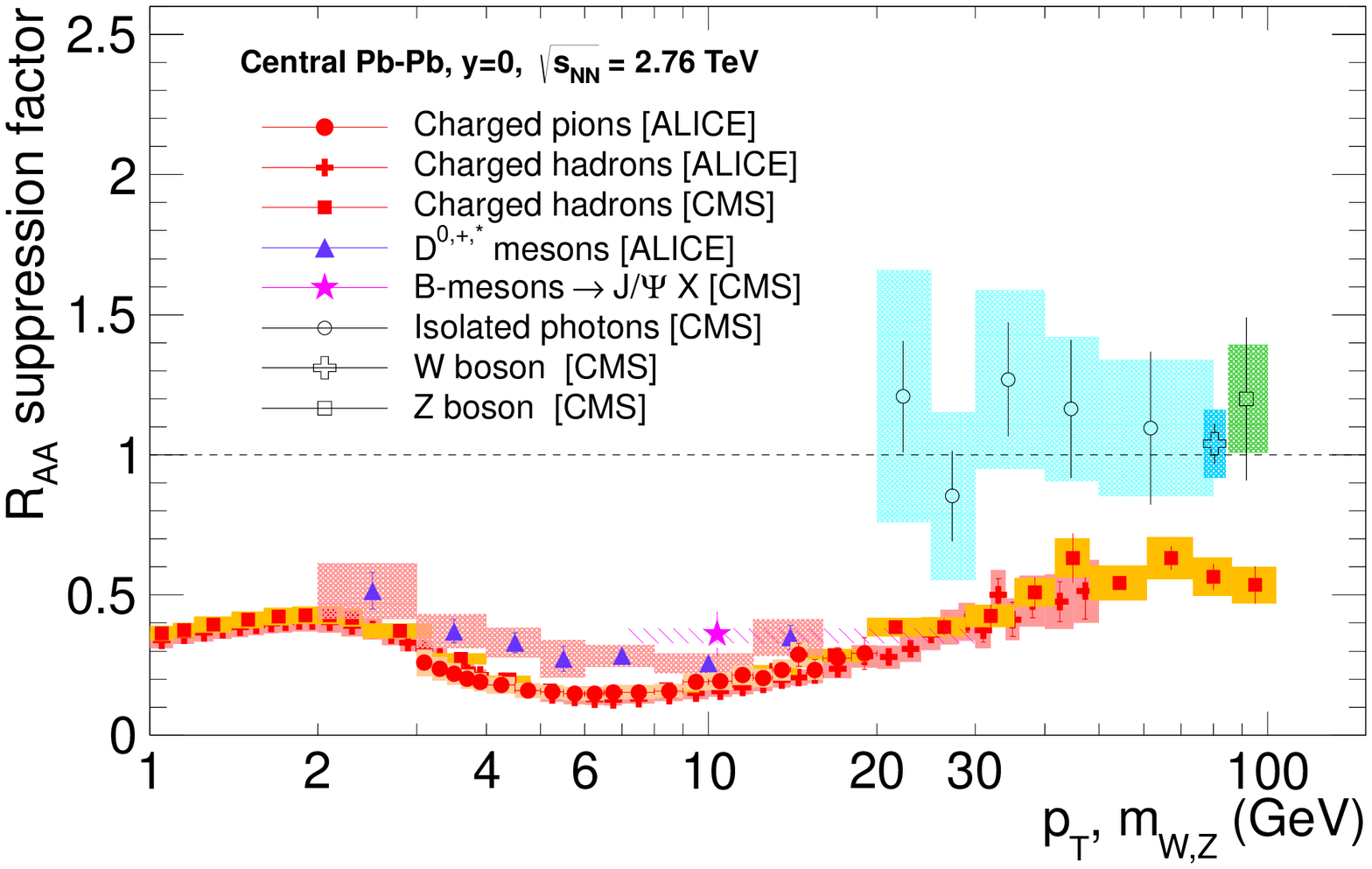}
\includegraphics[width=0.40\columnwidth,height=6.2cm]{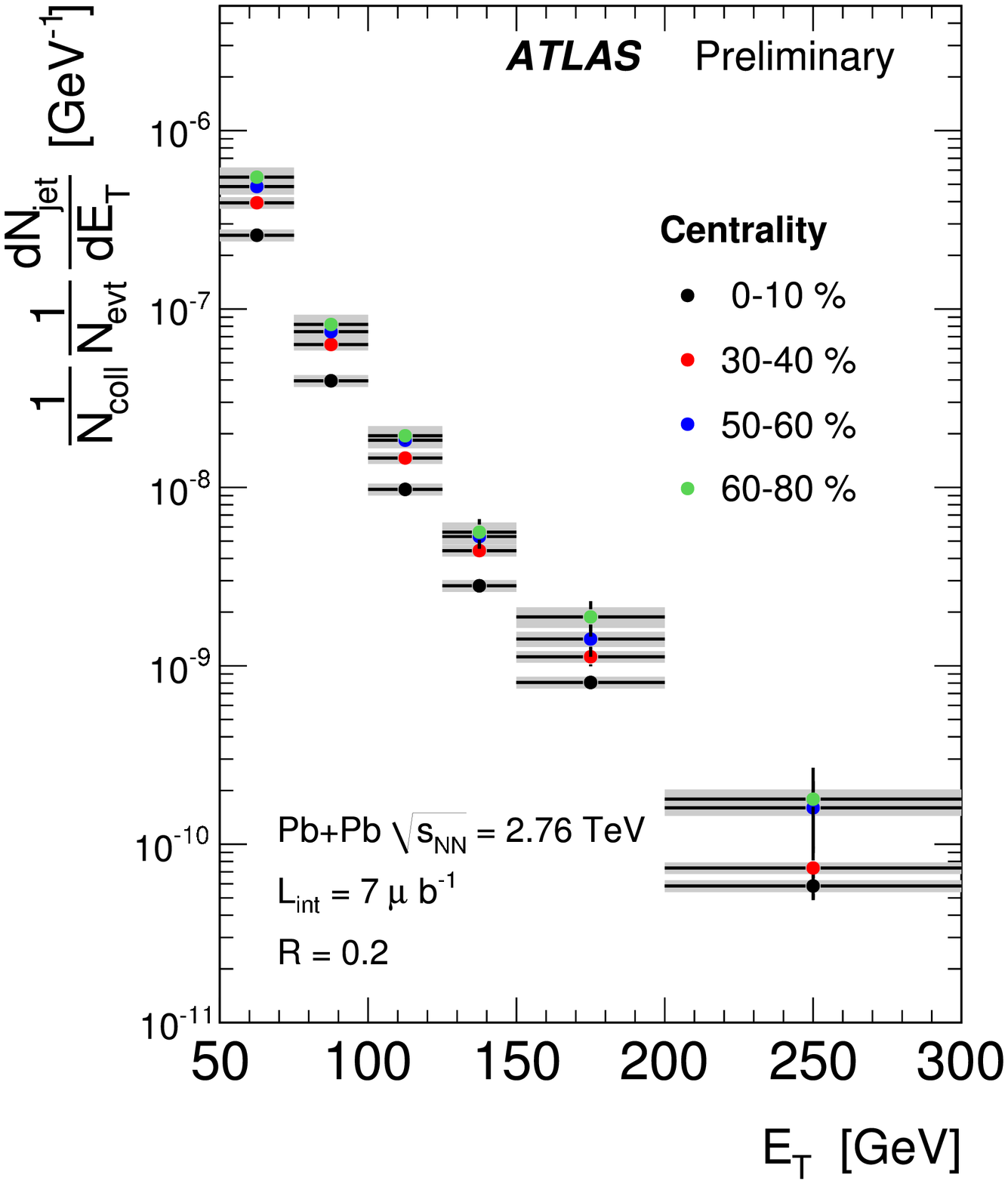}
\caption{Left: Compilation of measured $R_{AA}(p_{T})$ in central Pb-Pb collisions at the LHC 
for inclusive charged hadrons~\cite{Aamodt:2010jd,CMS:2012aa}, $\pi^\pm$~\cite{Appelshauser:2011ds}, 
D~\cite{ALICE:2012ab} and B~\cite{Chatrchyan:2012np} mesons, isolated-$\gamma$~\cite{Chatrchyan:2012vq},
and W~\cite{Chatrchyan:2012nt} and Z~\cite{Chatrchyan:2011ua} bosons. 
Right: Jet spectra (anti-$k_T$ algorithm, $R$~=~0.2) measured in various Pb-Pb
centralities scaled by their corresponding $T_{AA}$~\cite{atlas:jets12}.
The bands around all data points indicate the associated systematic uncertainties.} 
\label{fig:worldRAA}
\end{figure}

One of the first proposed smoking guns of QGP formation was ``jet quenching''~\cite{Bjorken:1982tu} 
i.e. the attenuation or disappearance of the spray of hadrons resulting from the fragmentation of a 
parton having lost energy in the dense plasma produced in the reaction.
Among the most exciting results from RHIC is the observation of large high-$p_T$ hadron suppression ($R_{AA}\approx$~0.2) 
in central Au-Au compared to p-p or d-Au collisions, in agreement with jet quenching 
expectations~\cite{Adcox:2004mh,Adams:2005dq}. At the LHC, charged hadrons appear to be further suppressed, 
by up to factor of 6 at $p_T$~=~7~GeV/c~\cite{Aamodt:2010jd,Appelshauser:2011ds} but the amount of suppression
is slowly reduced with increasing $p_T$ and plateaus at $R_{AA}\approx$~0.5 at
$p_T\approx$~40$-$100~GeV/c~\cite{CMS:2012aa} (Fig.~\ref{fig:worldRAA}, left). 
The observed $p_T$ dependence provides strong discrimination power for parton radiative energy loss models.
Once the underlying energy loss mechanism responsible for the observed high-$p_T$ hadron deficit is
confirmed~\cite{Milhano:2012gf} 
the properties of the medium (e.g. its $\qhat$ transport coefficient) can be properly derived 
taking into account the expansion of the produced matter in 3D-hydrodynamics calculations~\cite{Armesto:2009zi}.

\section{Jet quenching in dijet and $\gamma$-jet events}

\begin{figure}[htpb]
\includegraphics[width=5.cm,height=4.7cm, bb = 0 -30 916 700,clip=true]{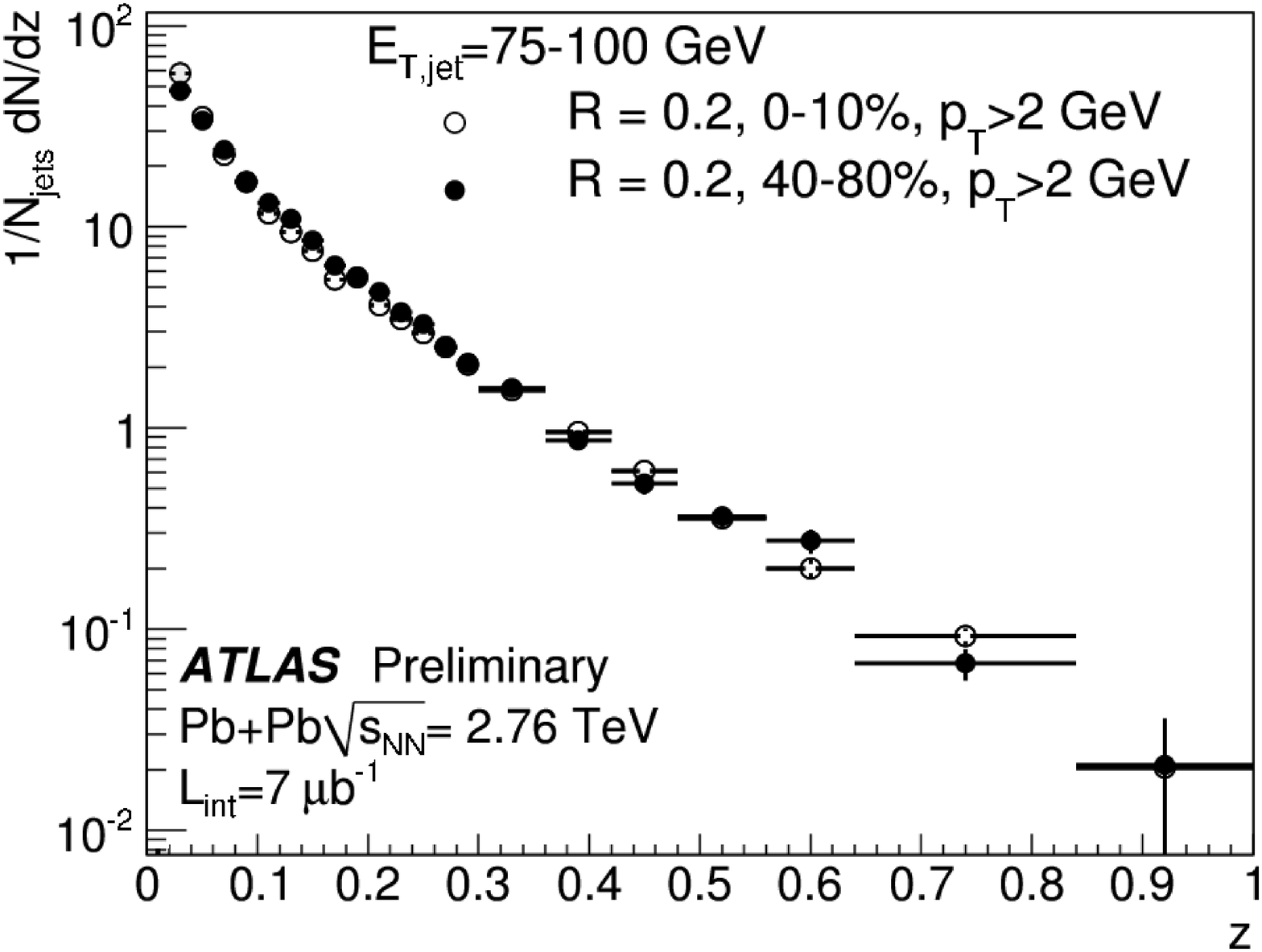}
\includegraphics[width=5.4cm,height=4.9cm]{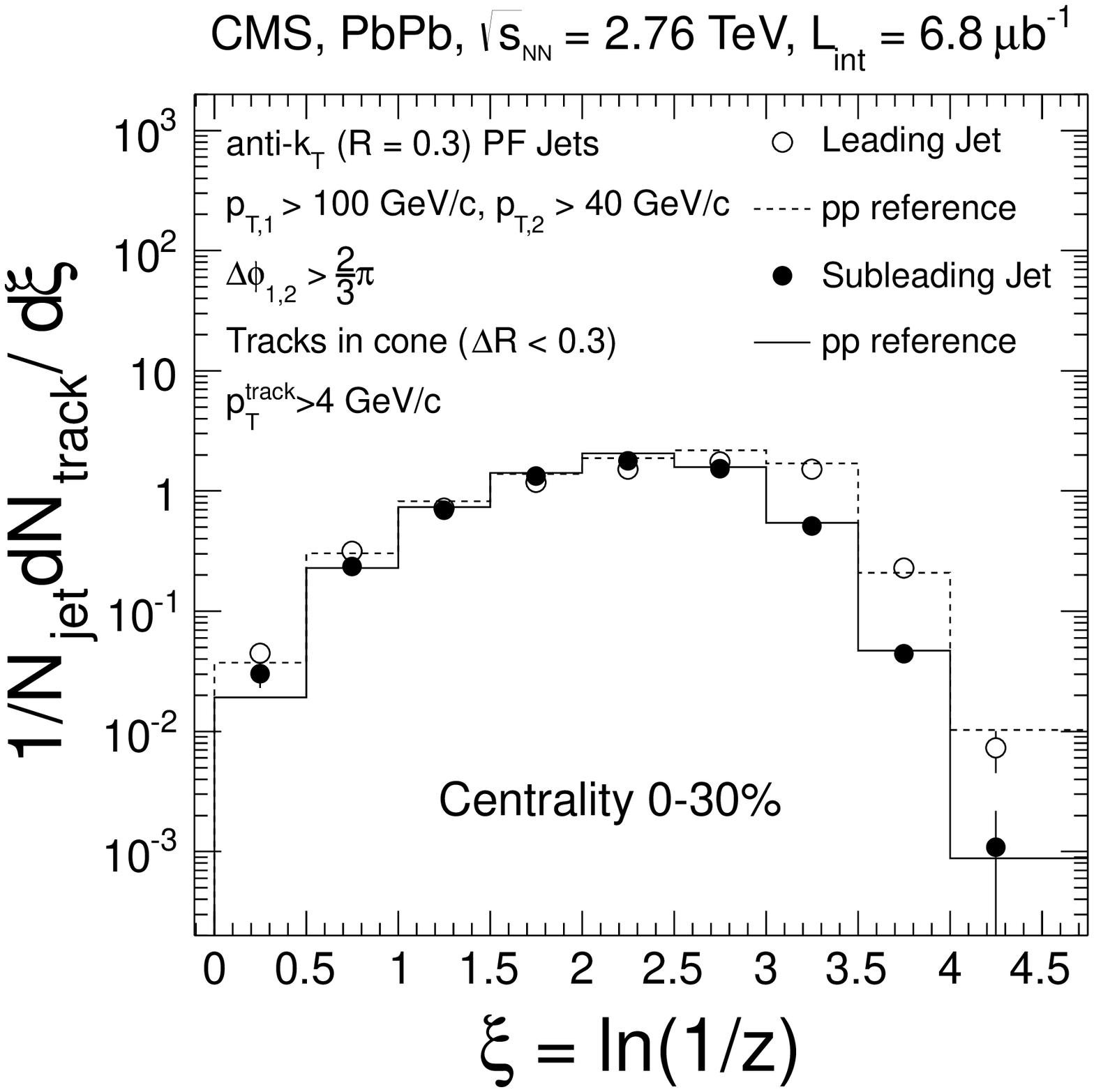}
\includegraphics[width=5.cm,height=4.5cm]{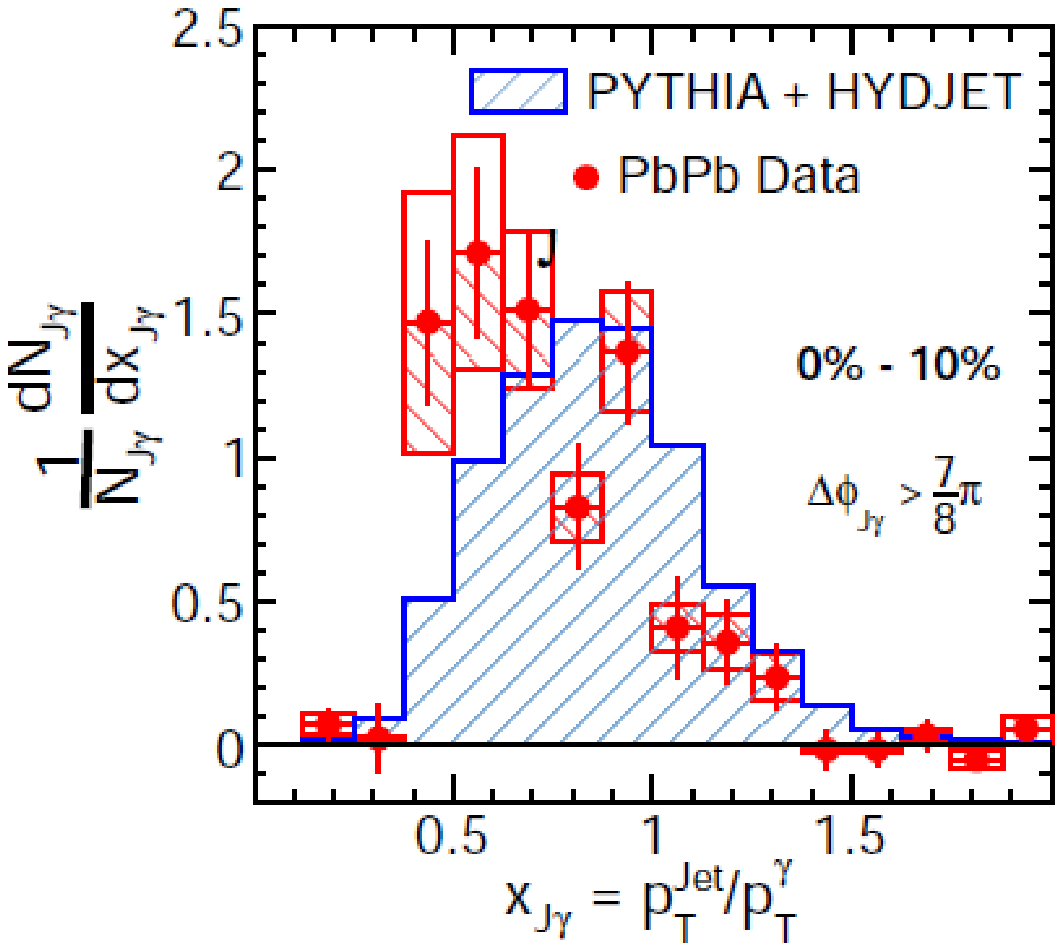}
\caption{Jet fragmentation functions reconstructed in Pb-Pb at 2.76 TeV as a function of the scaling variables
$z$ (left panel, for central and peripheral collisions)~\cite{atlas:jets12} and $\xi$ (mid panel, for central collisions)~\cite{Chatrchyan:2012gw}.
Right: Ratio of jet to photon momenta in central Pb-Pb collisions compared to {\sc pythia} p-p predictions~\cite{Chatrchyan:2012gt}.}
\label{fig:jets}
\end{figure}

High-$p_T$ jets have been fully reconstructed in heavy-ion collisions for the first time at the LHC 
using modern algorithms for jet finding and background subtraction~\cite{Cacciari:2010te}.
The strongest evidence so far for jet quenching is the observation of a large dijet momentum imbalance
for increasingly central Pb-Pb collisions~\cite{Aad:2010bu,Chatrchyan:2011sx} up to very high momenta 
($p_T\approx$~350~GeV/c)~\cite{Chatrchyan:2012nia}. In some cases, the quenched jet
is not even visible above the underlying event background (Fig.~\ref{fig:qgp_probes}, right).
As observed for the inclusive (leading) hadrons above 40~GeV/c, the jet spectra in central Pb-Pb are suppressed
by a factor of two compared to peripheral Pb-Pb collisions (Fig.~\ref{fig:worldRAA}, right)~\cite{atlas:jets12}.
The lost energy of the jet appears to be emitted in the form of soft particles ($p_T~<$~4~GeV/c) outside of
jet cone ($R >$~0.8)~\cite{Chatrchyan:2011sx}. Interestingly, even if the quenched jet loses a large
fraction of its energy, the dijet event keeps its back-to-back topology without significant azimuthal
decorrelation~\cite{Aad:2010bu,Chatrchyan:2011sx}. Fragmentation functions have been constructed using tracks
with $p_T~>$~2(4)~GeV/c within $R <$~0.2(0.3) of the jet-axis combined with the reconstructed (but quenched) 
jet energy~\cite{atlas:jets12,Chatrchyan:2012gw}.
The resulting FFs, shown in Fig.~\ref{fig:jets} as a function of the longitudinal fraction of the jet momentum
carried by the charged particles $z = p_{||}^{\rm track}/p^{\rm jet}$ (left) and of $\xi = \ln(1/z)$ (center), 
are similar in Pb-Pb (central and peripheral) and p-p collisions indicating that after energy loss the 
quenched jet fragments in the vacuum with the same pattern as ``normal'' jets.
Jet quenching has also been studied in ``cleaner'' photon-jet events
with $p_T^\gamma >$~60~GeV/c and $p_T^{\rm jet} >$~30~GeV/c~\cite{Chatrchyan:2012gt}.
The balance ratio $x_{\rm J\gamma} = p_T^{\rm jet}/p_T^\gamma$ is seen to decrease significantly 
with increasing centrality (see Fig.~\ref{fig:jets} right, for central Pb-Pb) as does the fraction of $\gamma$-jet
associations found in the data, compared to p-p and {\sc pythia}~\cite{Sjostrand:2006za} results.
Yet, the $\gamma$-jet azimuthal distribution measured in Pb-Pb events with large $p_T$ imbalance 
is consistent with that measured in p-p collisions and simulated with {\sc pythia}.
The fact that the quenched jet is back-to-back to the photon confirms that the energy 
is not lost in single hard gluon radiation.

\section{Heavy-flavour suppression}

In QCD, gluon-strahlung from heavy quarks is expected to be less important than for light quarks 
due to the ``dead-cone'' effect~\cite{Dokshitzer:2001zm}. Thus, a robust prediction of jet quenching models is
the hierarchy $\Delta E_{Q} < \Delta E_{q} < \Delta E_{g}$ accounting for mass and colour-charge dependencies
of radiative energy loss. Surprisingly, RHIC measurements of high-$p_T$ electrons from semileptonic $D$ and
$B$ decays seem to indicate a charm+bottom suppression comparable to that of light quarks: 
$R_{AA}(c,b)\approx R_{AA}(q,g)\approx$~0.2~\cite{Adare:2010de,Abelev:2006db}. At the LHC,  D-mesons have been for the
first time directly measured in Pb-Pb events with displaced-vertices in three hadronic decay channels
($D^0\to K\pi$, $D^\pm \to K\pi\pi$, $D^\star \to D^0\pi$ for $p_T >$~2,~4 and~6~GeV/c
respectively)~\cite{ALICE:2012ab}, and B-meson production has been observed via secondary $\jpsi$ for which a
clean separation of the displaced vertex is also possible for $p_T >$~6.5~GeV/c~\cite{Chatrchyan:2012np}. 
Within uncertainties, the measured suppression factors, $R_{AA}$(D)~$\approx$~0.3 and
$R_{AA}$(B)~$\approx$~0.4 are larger than $R_{AA}$(h$^\pm$,$\pi^\pm$)~$\approx$~0.15--0.2 
(Fig.~\ref{fig:worldRAA}, left) and support the expected radiative energy loss
hierarchy. Coming measurements with larger data samples and reduced uncertainties
will allow one to better understand the behaviour of heavy-quarks in the QGP, including the relative contribution 
of radiative/elastic losses, and determine e.g. their drag coefficients $\eta_{D}$ in the medium~\cite{Gossiaux:2012th}.

\section{Quarkonia dissociation}

\begin{figure}[htpb]
\includegraphics[width=0.55\columnwidth,height=6.2cm]{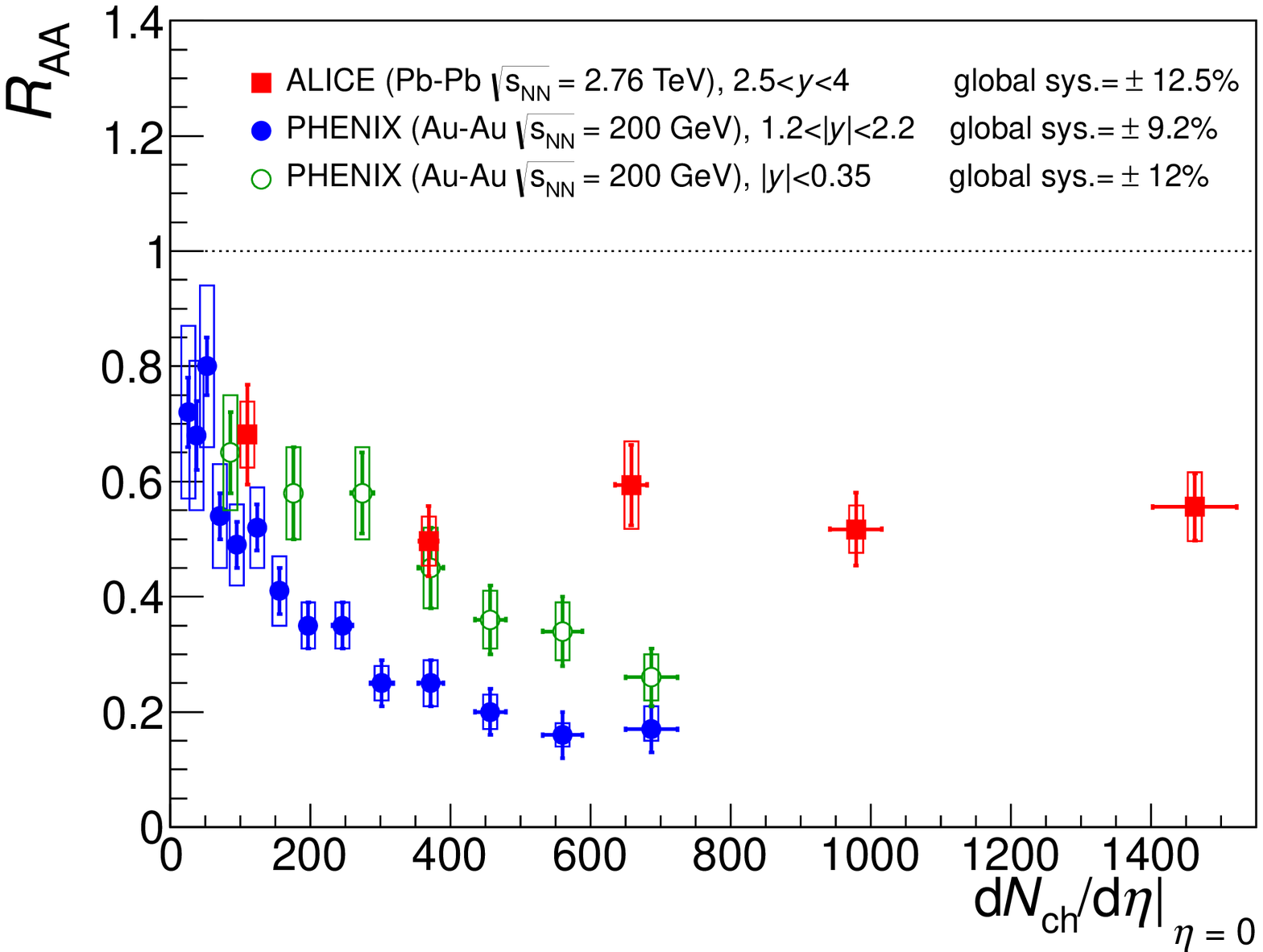}
\includegraphics[width=0.45\columnwidth,height=6.1cm]{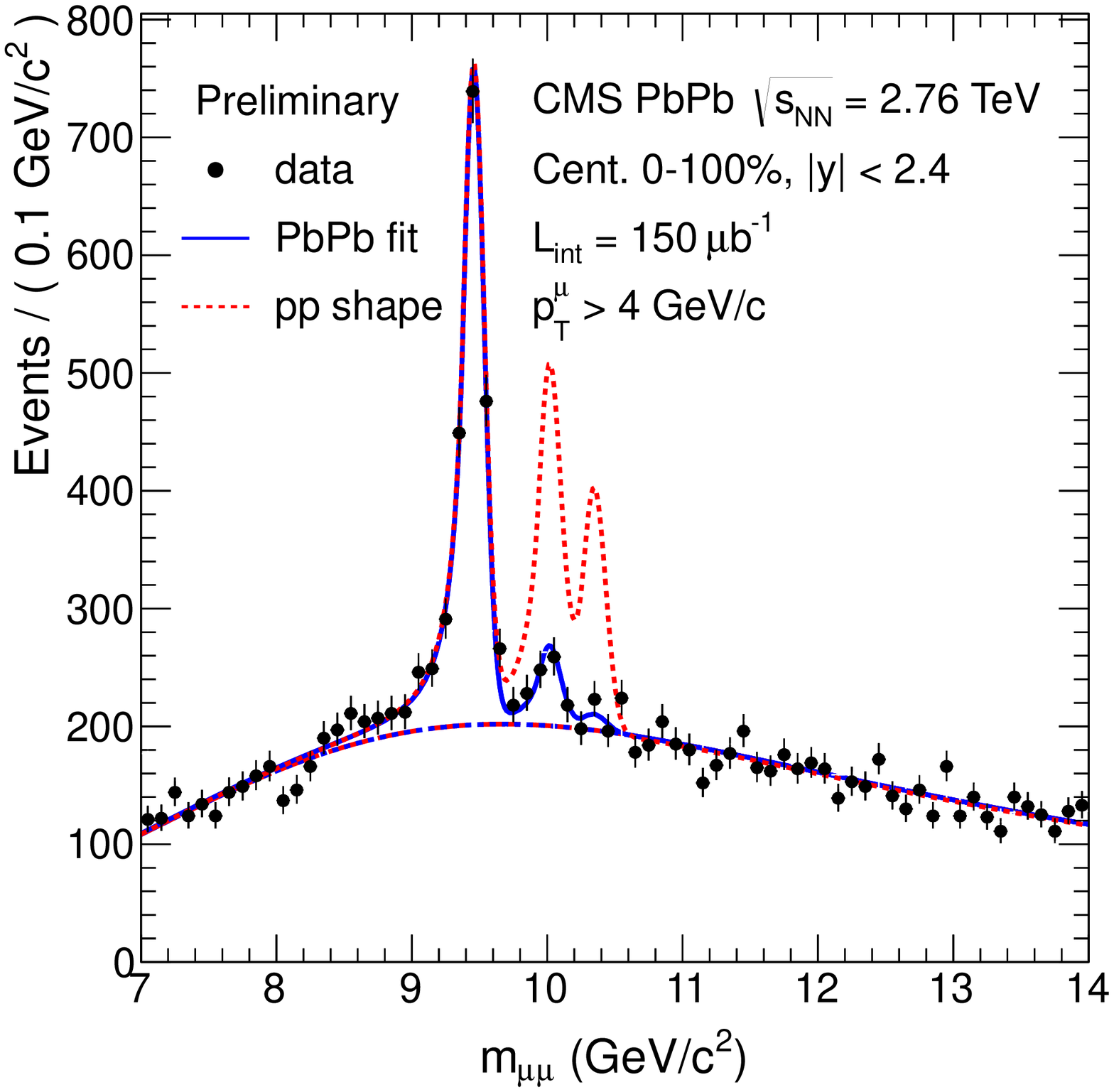}
\caption{Left: Inclusive $\jpsi$ $R_{AA}$ as a function of the central charged-particle density
in Pb-Pb at 2.76~TeV compared to PHENIX results in Au-Au at 200~GeV at central and forward rapidities~\cite{Abelev:2012rv}.
Right: $\Upsilon$(nS) states measured in Pb-Pb (data points) compared to the fitted 
p-p distribution (dashed red line) at 2.76~TeV~\cite{cms:Ups12}.}
\label{fig:jpsi_ups}
\end{figure}

The study of heavy-quark bound states in high-energy A-A collisions has 
a long story as a sensitive probe of the thermodynamical properties of deconfined QCD matter~\cite{Matsui:1986dk}. 
Analysis of quarkonia correlators and potentials in finite-$T$ lattice QCD~\cite{Datta:2003ww} indicate that the different 
$c$-$\bar c$ and $b$-$\bar b$ states dissociate at temperatures for which the colour (Debye) screening radius 
of the medium falls below their corresponding $Q\bar{Q}$ binding radius. 
The surprisingly similar amount of $\jpsi$ suppression observed at SPS~\cite{Alessandro:2004ap} and
RHIC~\cite{Adare:2006ns,Adare:2011yf} energies (with expected larger QGP temperature in the latter) 
has been interpreted as due to a partial compensation at RHIC of the reduced $\jpsi$ yields by $c\bar{c}$
recombination in the medium~\cite{BraunMunzinger:2000px,Thews:2000rj}. In this scenario, the large charm pair production 
expected at the LHC would further enhance charmonium regeneration~\cite{Andronic:2007bi}. 
LHC data on $\jpsi$ production~\cite{Aad:2010aa,Chatrchyan:2012np,Abelev:2012rv} show a different 
suppression pattern than observed at RHIC (Fig.~\ref{fig:jpsi_ups}, left). At low $p_T$, $\jpsi$ are less
suppressed ($R_{AA}\approx$~0.5) than measured at lower c.m. energies ($R_{AA}\approx$~0.2--0.3), 
whereas at high-$p_T$ they are more suppressed ($R_{AA}\approx$~0.2) than at RHIC ($R_{AA}\approx$~0.6).
Approaches based on deconfinement followed by charm recombination can reproduce the observed trends in the data 
although the model parameters ($\sigma_{c\bar c}$, density, ...) need to be validated with other LHC observations.
The abundant production of the $\ups(1S,2S,3S)$ states at the LHC opens up a unique opportunity
to study also the dissociation of the $b$-quark bound states. The $\ups$ is expected
to survive up to $4\,\Tcrit$ whereas the less tightly-bound $\upsp$ and $\upspp$ resonances should ``melt'' at lower temperatures.
The data confirms that the $\ups$ states are suppressed in central Pb-Pb relative to p-p
collisions: 
$R_{AA}(\ups) \approx$~0.56, $R_{AA}(\upsp) \approx$~0.12 and $R_{AA}(\upspp) <$~0.1
(Fig.~\ref{fig:jpsi_ups}, right)~\cite{Chatrchyan:2011pe,cms:Ups12}. 
The deficit in the $\ups$ production yields coincides with the expected fraction from the feed-down
contributions of the heavily-suppressed $\ups(2S,3S)$ excited states.  
The overall scenario is consistent with sequential suppression of the bottomonium family.

\section{Summary}

The main findings in the perturbative sector of Pb-Pb collisions at $\sqrtsnn$~=~2.76~TeV 
after 2 years of operation at the LHC can be summarised as follows:
\begin{itemize}
\item Electroweak probes ($\gamma$, W, Z) unaffected by final-state interactions can provide, with larger
  datasets and reduced uncertainties, valuable constraints of the nuclear PDFs~\cite{Chatrchyan:2012vq,Aad:2010aa,Chatrchyan:2011ua,atlas:Z12,Chatrchyan:2012nt}.
\item High-$p_T$ (leading) hadron production is suppressed compared to p-p collisions by factors ranging from 6
  (at 7~GeV/c) to 2 (above 40~GeV/c), indicating the formation of a dense opaque medium that suppresses the
  energy of the parent fragmenting partons~\cite{Aamodt:2010jd,Appelshauser:2011ds,CMS:2012aa}.
\item Fully reconstructed jets are quenched in dijet and $\gamma$-jet events and show 
  (i) a large $p_T$ imbalance, (ii) enhanced soft off-cone radiation, 
  (iii) preserved back-to-back azimuthal correlations, and (iv) vacuum-like fragmentation
  functions (reconstructed using the quenched jet energy)~\cite{Aad:2010bu,Chatrchyan:2012gt,atlas:jets12,Chatrchyan:2011sx,Chatrchyan:2012nia,Chatrchyan:2012gw},
  which suggest that parton energy loss occurs via soft multi-gluon emission followed by 
  vacuum fragmentation.
\item D and B mesons are suppressed by factors of $\sim$4 and $\sim$3 respectively relative to p-p
  collisions~\cite{ALICE:2012ab,Chatrchyan:2012np}, in agreement with the expected mass and colour-charge dependencies of radiative energy loss.
\item The deficit of $\jpsi$ yields at the LHC is smaller than at lower energies. The suppression is weaker
  (stronger) at low (high) $p_T$~\cite{Aad:2010aa,Chatrchyan:2012np,Abelev:2012rv} and is
  suggestive of model predictions based on colour screening followed by $c\bar c$ recombination.
\item 
  The $\upsp$ and $\upspp$ states are strongly suppressed, whereas the $\ups$ ground-state yield is depleted by about 
  40\% (the same fraction expected from feed-down contribution of the two excited states) relative to p-p collisions~\cite{Chatrchyan:2011pe,cms:Ups12},
  in agreement with sequential dissociation scenarios.
\end{itemize}
Figure~\ref{fig:worldRAA} (left) compiles the $R_{AA}(p_T)$ factors for most of the
high-$p_T$ and large-mass particles measured in central Pb-Pb collisions at the LHC. 
The mechanisms of initial production of all particles are well understood within pQCD. 
The final reduced yields observed for all particles containing colour degrees of freedom 
clearly point to strong final-state effects (radiative energy loss, colour screening, ...).
Final extraction of the thermodynamical and transport properties 
of the hot and dense matter produced at the LHC will require detailed calculations including mechanisms
of particle production+``destruction'' coupled with a (3D viscous) hydrodynamics 
description of the expansion of the system.\\

\hspace{-0.65cm}{\it \bf Acknowledgments:} I am grateful to Bernard Pire and Michel Gar\c{c}on for their invitation to 
this well-organised and interesting multidisciplinary conference.

\end{document}